\documentclass{kapproc} 

\usepackage{procps} 
\usepackage[dvips]{graphicx}

\upperandlowercase

\kluwerbib 

\begin{document}

\articletitle{The Lenses Structure \& Dynamics Survey}

\articlesubtitle{The internal structure and evolution of E/S0 galaxies
and the determination of H$_0$ from time-delay systems}

\author{L.V.E. Koopmans$^{1,2}$ \& T.Treu$^2$}
\affil{$^1$Space Telescope Science Institute,
3700 San Martin Drive, Baltimore, MD 21218, USA}
\affil{$^2$California Institute of Technology, mailcode 105--24, Pasadena,
CA 91125, USA}
\email{koopmans@stsci.edu, tt@astro.caltech.edu}


\begin{abstract}
The Lenses Structure \& Dynamics (LSD) Survey aims at studying the
internal structure of luminous and dark matter -- as well as their
evolution -- of field early-type (E/SO) galaxies to $z\sim1$. In
particular, E/S0 lens galaxies are studied by combining gravitational
lensing, photometric and kinematic data obtained with ground-based
(VLA/Keck/VLT) and space-based telescopes (HST). Here, we report on
preliminary results from the LSD Survey, in particular on (i) the
constraints set on the luminous and dark-matter distributions in the
inner several $R_{\rm eff}$ of E/S0 galaxies, (ii) the evolution of
their stellar component and (iii) the constraints set on the value of
H$_0$ from time-delay systems by combining lensing and kinematic data
to break degeneracies in gravitational-lens models.
\end{abstract}


\section*{Introduction}

Even though massive E/S0 galaxies enclose most mass (luminous and
dark) in the Universe on galactic scales, relatively little is
observationally known about their their dark-matter halos or the
evolution of their internal structure with time, and only recently
studies have started to shed some light on the evolution of their
stellar population with redshift.

The reasons for this are both observational and intrinsic to the mass
modeling. First, since many of the studies of the mass distribution of
E/S0 galaxies rely on stellar kinematics through spectroscopic
observations, only with the advent of 8-10\,m class telescopes
(e.g. Keck and VLT) has it becomes possible to study E/S0 galaxies in
any detail beyond the local Universe. Second, degeneracies in mass
models that rely only on kinematic and photometric data often allow
for multiple solutions and place limited or no constraints on the
presence and distribution of dark-matter. These problems exacerbate
with increasing redshift due to poorer observational constraints.  Model
degeneracies are often due to the unknown mass of the galaxy, allowing
one to freely play with stellar anisotropy, the mass-density slope and
its normalization. 
For example, approximately constant velocity
dispersion profiles can be explained with an isothermal, kinematically
isotropic, luminous plus dark-matter distribution, as well as a
constant stellar $M/L$ model with a radially increasing tangential
anisotropy. However, the latter model requires significantly less mass
than the isothermal model within, say, several effective radii
($R_{\rm eff}$).

Hence, if the total mass of an E/S0 galaxy enclosed within some radius
(around $\sim R_{\rm eff}$) is known, one can break the degeneracy
between the stellar $M/L$, the stellar anisotropy and the radial mass
distribution. Strong gravitational lensing by E/S0 galaxies provides
exactly the required information!

\section*{The Lenses Structure \& Dynamics (LSD) Survey}

The `clean' LSD Survey sample consists of 11 relatively isolated
(e.g. no massive clusters nearby) E/S0 lens galaxies spread between
redshifts $z=0.04$ and 1.01. The galaxies have a mass range of
$\sim$1.5 orders of magnitude. Multi-color photometric data is
available in the HST archive (mostly from the CASTLeS collaboration)
for each system (typically V, I and H bands). In 2001--2002, we
obtained stellar kinematic data using the Echelle Spectrograph and
Imager (ESI) on the Keck--II telescope with typically 0.7$''$ seeing
and under photometric conditions. Some systems have extended kinematic
profiles (along major and sometimes minor axes), others only
luminosity weighted dispersions, depending on their brightness and
extent. Besides the clean sample, we also observed several other
systems (also with the VLT), including several disk-galaxy lenses and
lens systems with measured time-delays.

We continue with some of the high-lights of the LSD survey and
related studies (e.g. the determination of H$_0$ from lensing).

\section*{The evolution of the stellar mass-to-light ratio}

Two LSD systems (MG2016+112 and 0047--285) have thus far been studied
in detail (Koopmans \& Treu 2002, 2003; Treu \& Koopmans 2002a). Since
we have available the (central) stellar velocity dispersion, effective
radius and effective surface brightness (from the HST images;
transformed to rest-frame B-band), each lens system can directly be
compared with the local Fundamental Plane (FP). The offset from the FP
is an indicator of the evolution of the effective surface brightness,
due to fading of the stellar population with time (i.e. ``passive
evolution''). Both systems are consistent with passive evolution of
field E/S0 galaxies, marginally faster than that of cluster E/S0
galaxies. We find that the stellar $M/L$ determined from the FP
evolution and local measurements, being close to the ``maximum-bulge
solution'', agree with those measured {\sl only} from lensing and
dynamics, suggesting that no structural evolution has occurred in the
FP below $z\sim 1$.

\section*{The luminous and dark-matter mass profile}

The combination of stellar kinematics and gravitational lensing can
also be used to place stringent constraints on the
luminous plus dark-matter mass profiles of E/S0 galaxies to $z\sim1$.
The reason is that lensing determines the mass inside the Einstein
radius to a few percent accuracy. Varying the inner (total) mass
slope -- but satisfying the stringent mass constraint -- leads to a
considerable change in the observed line-of-sight stellar velocity
dispersion profile, as well as the luminosity weighted stellar
dispersion, of the E/S0 galaxy, only weakly dependent on details such
as anisotropy, etc (see KT). Thence, a comparison with the observed
stellar kinematics allows the determination of an `effective' slope
($\gamma'$ for $\rho\propto r^{-\gamma'}$; i.e. the average luminous
dark-matter power-law slope) inside the Einstein radius (typically
1--5\,$R_{\rm eff}$).

Thus far, we have found that $\gamma'=2.0\pm0.1\pm0.1$ for MG2016+112
(Treu \& Koopmans 2002a) and $\gamma'=1.9^{+0.05}_{-0.23}\pm0.1$ for
0047$-$285 (Koopmans \& Treu 2003; 68\% C.L. and syst. error). For
B1608+656--G1 and PG1115+080, both not part of the `clean' LSD sample,
we find $\gamma'=2.03\pm 0.14 \pm 0.03$ and $\gamma'=2.35\pm 0.1 \pm
0.05$, respectively (Koopmans et al. 2003; Treu \& Koopmans
2002b). E/S0 galaxies in the clean sample are both consistent with
isothermal mass profiles (i.e.~$\gamma'=2$). From all four systems
studied thus far, the average is $\langle \gamma'\rangle = 2.1$ with
an rms of 0.2; {\sl E/S0 galaxies appear on-average isothermal to
within $\sim$ 10\% inside several effective radii, eventhough some
intrinsic scatter between systems is found, as expected.}

We note that this is a preliminary results and the analysis of the
full sample is required to confirm/strengthen this conclusion.  Even
so, E/S0 galaxies require a considerable diffuse dark-matter component
inside the stellar spheroid in order to explain the observed stellar
kinematics; constant stellar $M/L$ models are excluded with high
confidence in all cases. The luminous plus dark-matter appears to
conspire to form an isothermal profile in its inner regions, similar
to that observed for spiral galaxies.  Finally, upper limits have been
set on the inner dark-matter profile of E/S0 as well as the
(an)isotropy of the stellar component, but we defer a discussion until
the entire sample has been analyzed.

\section*{The value of H$_0$ from lens time-delays}

The most severe degeneracy in lens models is that of the (unknown)
slope of the radial mass profile of the lens mass distribution.
Different power-law slopes (other than e.g. isothermal) can often
equally well fit the same lensing-only constraints (e.g. Wucknitz
2002). Different slopes, however, lead to different inferred values of
H$_0$ from time delays, roughly following $\Delta {\rm H}_0/{\rm
H}^{\gamma'=2}_0 = (\gamma'-2)$; i.e. steeper (shallower) than
isothermal mass profile lead to higher (lower) inferred values of
H$_0$. 

Since the combination of stellar kinematics and gravitational lensing
can tightly constrain $\gamma'$ (see above; assuming the mass profile
indeed follows approximately a power-law), the usefulness of this to
time-delay lenses and the determination of H$_0$ is apparent. We have
thus far looked at two systems in detail and find
H$_0=59^{+12}_{-7}\pm3$~km~s$^{-1}$~Mpc$^{-1}$ (PG1115+080) and
H$_0=75^{+7}_{-6}\pm4$~km~s$^{-1}$~Mpc$^{-1}$ (B1608+656) for
$(\Omega_{\rm m},\Omega_{\Lambda})=(0.3,0.7)$. The errors are the 68\%
C.L. and systematic errors and include a realistic uncertainty due to
the slope of the radial mass profile.  These values should therefore
be relatively unbiased. The deviation of PG1115+080 from isothermal
(see above) increases H$_0$ by $\sim$35\% from 44 to
59~km~s$^{-1}$~Mpc$^{-1}$, exemplifying the need to measure the
stellar kinematics of lens galaxies for each time-delay system. Both
values are consistent with local determinations of H$_0$
(e.g. Freedman et al. 2001), but inconsistent with the sample studied
by e.g. Kochanek \& Schechter (2003; and references therein).  We are
currently observing more systems with Keck and the VLT to improve the
statistics of the sample and further examine the nature of this
apparent inconsistency between lens systems.

\section*{Conclusions}

The combination of stellar kinematic with gravitational lensing
provides a powerful new tool to study the internal structure and
evolution of E/S0 galaxies. Some of the preliminary results from the
LSD Survey indicate that E/S0 lens galaxies to $z\sim 1$ evolve
passively and have nearly-isothermal luminous plus dark matter mass
profiles inside several R$_{\rm eff}$. Application of this to lens
systems with time-delays gives more accurate values of H$_0$, sofar in
agreement with local determinations. The study of more lens systems is
required to confirm/strengthen these conclusion, but results so far
have been encouraging.

\end{document}